\documentclass[12pt,preprint]{aastex}

\slugcomment{DRAFT}

\shorttitle{Characterizing the AO Off-Axis PSF - I}
\shortauthors{Steinbring et al.}

\input epsf
\def\plotone#1{\centering \leavevmode
\epsfxsize=1.0\columnwidth \epsfbox{#1}}

\def\plotonehalf#1{\centering \leavevmode
\epsfxsize=0.5\columnwidth \epsfbox{#1}}

\def\plottwo#1#2{\centering \leavevmode
\epsfxsize=0.5\columnwidth \epsfbox{#1}
\epsfxsize=0.5\columnwidth \epsfbox{#2}}

\begin{document}

\title{Characterizing the Adaptive Optics Off-Axis\\
Point-Spread Function - I:\\ 
A Semi-Empirical Method for Use in Natural-Guide-Star Observations\altaffilmark{1,2}}

\author{E. Steinbring\altaffilmark{3,4},
S. M. Faber\altaffilmark{5}, S. Hinkley\altaffilmark{6},
B. A. Macintosh\altaffilmark{7}, D. Gavel\altaffilmark{7},
E. L. Gates\altaffilmark{8}, Julian C. Christou\altaffilmark{3},
M. Le Louarn\altaffilmark{3,9}, 
L. M. Raschke\altaffilmark{5}, Scott A. Severson\altaffilmark{5}, 
F. Rigaut\altaffilmark{10}, David Crampton\altaffilmark{11,4}, 
J. P. Lloyd\altaffilmark{12}, and James R. Graham\altaffilmark{12}}

\altaffiltext{1}{Based, in part, on observations obtained at the Lick Observatory, which is operated by the University of California.}

\altaffiltext{2}{All authors except DC are affiliated with the Center for
Adaptive Optics.}

\altaffiltext{3}{Center for Adaptive Optics, University of California,
Santa Cruz, CA 95064}

\altaffiltext{4}{Visiting Astronomer, Canada-France-Hawaii Telescope,
operated by the National Research Council of Canada, the Centre de la
Recherche Scientifique de France, and the University of Hawaii.}

\altaffiltext{5}{UCO/Lick Observatory, Department of Astronomy and Astrophysics, University of California, Santa Cruz, CA 95064}
 
\altaffiltext{6}{Department of Physics, University of California, Santa Cruz, CA
95064. Current address: American Museum of Natural History, Department of Astrophysics, New York, NY 10024-5192} 

\altaffiltext{7}{Lawrence Livermore National Laboratory, P.O. Box 808, 
Livermore, CA 94551}

\altaffiltext{8}{Lick Observatory, P.O. Box 85, Mt. Hamilton, CA 95140}

\altaffiltext{9}{Current address: European Southern Observatory, Karl-Schwarzschild Str. 2, 85748 Garching bei M{\"u}nchen, Germany}

\altaffiltext{10}{Canada-France-Hawaii Telescope Corporation, Kamuela, HI 96743, Current address: Gemini Observatory, 670 N. A'ohoku Place, Hilo, HI 96720}

\altaffiltext{11}{Herzberg Institute of Astrophysics, National Research Council of Canada, Victoria, BC V9E 2E7, Canada}

\altaffiltext{12}{Department of Astronomy, University of California, Berkeley, CA 94720}

\begin{abstract} 
Even though the technology of adaptive optics (AO) is rapidly maturing, calibration of 
the resulting images remains a major challenge.
The AO point-spread function (PSF) changes quickly both in time and
position on the sky. In a typical observation the star used for guiding
will be separated from the scientific target by 10{\arcsec} to 30{\arcsec}. 
This is sufficient separation to render images of the 
guide star by themselves nearly useless in characterizing the PSF at the off-axis target position.  
A semi-empirical technique is described that improves the determination
of the AO off-axis PSF. The method uses calibration images of dense
star fields to determine the change in PSF with field position.
It then uses this information to correct contemporaneous images of the guide star to produce a PSF that is more accurate
for both the target position and the time of a scientific observation.
We report on tests of the method using natural-guide-star AO systems on the Canada-France-Hawaii Telescope and Lick Observatory Shane Telescope,
augmented by simple atmospheric computer simulations. At 25{\arcsec}
off-axis, predicting the PSF full width at half maximum using only information about the guide star results in an error of 60\%. Using
an image of a dense star field lowers this error to 33\%, and our
method, which also folds in information about the on-axis PSF, further
decreases the error to 19{\%}.

\end{abstract}


\section{Introduction}\label{introduction}

A key ingredient to successful analysis of images produced with the aid of an
adaptive optics (AO) system is the determination of the point-spread function
(PSF). The application dictates how precisely the PSF must be determined.
For example, limiting the uncertainty in PSF-fitting photometry in a crowded
star field to only a few percent will demand very high accuracy in knowledge of the Strehl ratio.
Another example, but a case where a good reference star is less 
likely to be found, is the subtraction of the point-like
core from the image of a quasar-host or radio galaxy
\citep{Hutchings1998, Hutchings1999, Steinbring2002}. In this case
Strehl ratio may be low, and the AO PSF approximately Gaussian. 
For a fixed volume, the peak of a 2-dimensional circular Gaussian is inversely
proportional to the square of its full-width at half maximum (FWHM). 
Therefore, reducing the FWHM by, say, 25\% will drive the peak up by 78\%. 
A 25\% underestimate of the QSO core FWHM will have a
detrimental effect on the quality of fit. In practice, accuracy of only 10{\%} to 20{\%} in PSF FWHM is sufficient for detection of the host galaxy. 

\begin{figure}
\plotonehalf{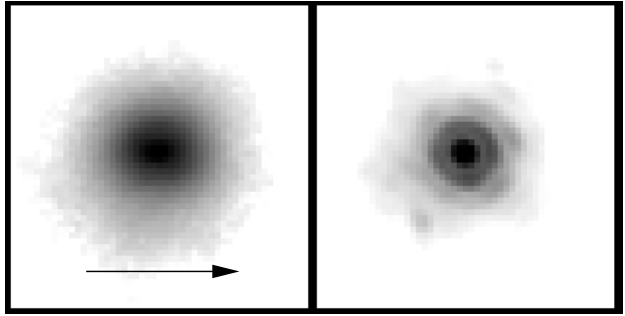}
\caption{Simultaneous AO observations of two well-separated stars.
The image of the guide star is shown on the right, and 
the image of a star at an offset of 20{\arcsec} is shown on the left. Both images are displayed with the same log grey-scale, and have a FOV of roughly
2{\arcsec} $\times$ 2{\arcsec}. The direction to the guide star is
indicated by the arrow in the left-hand panel. The effects of anisoplanatism
and anisotropy have degraded the off-axis PSF and elongated it towards the guide star.}
\label{figure_star_onaxis_offaxis}
\end{figure}

The task of determining the PSF is made difficult by three factors. First, the 
performance of the AO system depends on the brightness of the star which it
uses to measure correction - the ``guide star.''
Second, the delivered
correction depends directly on quickly changing seeing conditions. 
Third, the AO PSF is strongly field dependent (anisoplanatic).
The first factor requires that scientific observations
and calibration measurements employ either the same guide star or ones
of similar brightness. Choosing a fainter guide star for calibration may 
require, to preserve $S/N$, that the AO system obtain longer integration times per measurement. Phase corrugations change continuously
as the wind moves turbulence across the telescope pupil, and therefore
the lower bandwidth will result in poorer correction.
The latter two factors are related to the degree of turbulence along the
line of sight. The seeing scales linearly with the inverse of the
coherence length of phase distortions, 
$r_0$ (Fried's parameter), which for observations at zenith angle $\gamma$, 
are related to the atmospheric refractive-index structure ``constant'', $C_n^2$, and wavelength, $\lambda$, by
$$r_0(\lambda, \gamma)=0.185\lambda^{6/5}\cos^{3/5}{\gamma}
\Big(\int{C_n^2(h) dh}\Big)^{-3/5}, \eqno (1)$$
where $h$ is height in the atmosphere (for review papers see 
Roddier 1981, 1999; Beckers 1983). Thus a measurement of the PSF needs to be
simultaneous with the target observation or at least sample the same 
variation in $r_0$.
Differences in phase
distortion along different lines of sight, exacerbated by a $C_n^2(h)$ profile
skewed to high altitude, will dramatically degrade both Strehl ratio and
FWHM with increasing telescope offset from the guide star. A separation of as little as 10{\arcsec} is sufficient to render
images of the guide star unusable as a PSF estimate for the target.
The effect of tilt correlation also causes the off-axis PSF to be anisotropic.  For any significant offset the amount of beam overlap between guide and PSF 
star in the tangential direction (perpendicular to the offset) will be greater than for the sagittal direction (along the offset). The lower correlation of aberrations
in the sagittal direction causes the PSF to be elongated towards the guide star (see e.g. McClure et al. 1991,  Voitsekhovich \& Bara 1999).  Limited AO system bandwidth may also elongate the PSF in the direction of the prevailing wind.
Thus the PSF can depend on position angle in the frame, and
this requires that the PSF measurement be made near the target
position as well.
Figure~\ref{figure_star_onaxis_offaxis} illustrates the dramatic spatial dependence of the PSF. It shows
natural-guide-star AO observations of two well-separated stars,
taken simultaneously  (these data are discussed later, in Section \ref{cfht_pueo}).
The right-hand panel is an image of the guide star and the left-hand
panel is an image of a star at an offset of 20{\arcsec}. Both images have 
a field of view (FOV) of roughly
2{\arcsec} $\times$ 2{\arcsec}, and the direction to the guide star is 
indicated by the arrow in the left-hand panel. 
The on-axis image has a high Strehl ratio, greater than 30\%, but the
off-axis Strehl ratio is closer to 10\%. 
Note the
degradation of the off-axis PSF and its elongation towards the guide star.

If a sufficiently bright star were within a few arcseconds of
the target, the task of PSF measurement would be greatly simplified. The observer could use
this ``PSF star'' as a nearby and simultaneous measurement of the PSF during the
target observation. Unfortunately, the probability that a suitably bright
PSF reference will be available is very small. 
Furthermore, some AO systems 
operate with imagers originally designed for uncompensated seeing, and their new
optics have sacrificed field of view in order to obtain Nyquist
sampling. For example, the small FOV such as the 4{\arcsec} $\times$ 4{\arcsec} provided by the $256\times256$ pixel Keck AO Near-Infrared Spectrometer (NIRSPEC) imager makes a fortuitous observation of both target and PSF star even more unlikely.

When faint extended objects are observed with natural guide-star AO, they will 
always be offset from the guide star. For a large FOV imager such as the Canada-France-Hawaii Telescope (CFHT) $1024\times1024$ pixel infrared camera (KIR), the guide star is often within the 36{\arcsec} $\times$ 36{\arcsec} FOV of the detector. This large FOV has the advantage of ensuring that a bright 
star will be available to characterize the on-axis PSF. Unfortunately, target integrations are typically long enough to leave a saturated image of
the star. Still, several very short
calibration exposures could be interleaved
with those observations. An average of these would provide
an unsaturated image of the star to characterize the on-axis PSF.

But even without images of the guide star, one can still determine the 
on-axis PSF.
Consider a wave-front with corrugations consistent with a Kolmogorov spectrum. Now, consider perfect correction of the wave-front phase over an unobstructed pupil. An analytic expression can 
be found that gives the improvement in Strehl ratio due to successive
removal of an increasing number of Zernike polynomials from the aberrated wave-front \citep{Noll1976}. For example, removing the
first 3 Zernike modes (excluding piston) corresponds to a reduction in the residual mean square wave-front distortion by a factor of 5. In a real
AO system, these residual errors can in principle be determined from wave-front sensor measurements and thus used to predict the delivered on-axis PSF. The wavelengths of sensing and correction may also be different. The light from the 
guide star is usually split; optical light 
is sent to the wave-front sensor while correction is applied to the instrument light path in the near-infrared.
\citet{Veran1997} have exploited this and developed software that uses 
wave-front information 
to predict the on-axis PSF in the near-infrared for the CFHT AO system.

Knowing the on-axis PSF will not give the PSF at the target 
position. However, an estimate of the change in the PSF with increasing 
offset can be determined theoretically.
For the case of a single thin turbulence layer at height $h$, infinite telescope aperture, and perfect correction of an infinite number of modes, the 
off-axis Strehl ratio, $S$, falls as \citep{Roddier1999}
$$S\simeq\exp[-(\theta/\theta_0)^{5/3}] \eqno (2)$$
where $\theta$ is the offset angle on the sky, and
$$\theta_0=0.314{{r_0 \cos{\gamma}}\over{h}} \eqno(3)$$
is the isoplanatic angle. For $r_0(0.5 \mu{\rm m})=10$ cm and $h=2$ km this gives $\theta_0=19${\arcsec}
for $K$-band observations at zenith. That is, the PSF at an offset
of 19{\arcsec} will have a Strehl ratio degraded by a factor of $e$
from the on-axis value.  

Determining the real AO off-axis PSF is more complicated.
A real $C_n^2(h)$ profile will contain significant power
at low altitude (dome seeing, ground-layer), but the more important 
contribution to anisoplanicity occurs at higher altitude. 
For example, measurements using SCIntillation Detection And Ranging (SCIDAR) show that a single thin high-altitude layer typically dominates the nighttime 
free-atmosphere $C_n^2(h)$ profile above Mauna Kea \citep{Racine1995}.
These observations predate the use of so-called generalized SCIDAR, however,
and thus do not give a true picture of the conditions at the ground layer. 
But, evidence from generalized SCIDAR measurements at other observatories suggest that dominance by at most a few high layers may occur elsewhere as well
(cf. Klueckers et al. 1998).
Equation 3 suggests that the isoplanatic angle is inversely proportional to 
the height of the dominant layer. In reality, the off-axis Strehl ratio
will depend on the mean height of the atmospheric turbulence, $\bar{h}$, and 
on the equivalent number of corrected Zernike modes, $N$. Higher order
modes will decorrelate more quickly than low order ones, resulting in 
a smaller isoplanatic angle (see e.g. Roddier 1999).
That is, equation 2 should be replaced by \citep{Flicker2002}
$$S\simeq\exp[-(\theta/\theta_N)^{5/3}], \eqno (4)$$
with the exponent tending towards 2 for very partial compensation \citep{Roddier1993}. The ``instrumental'' isoplanatic angle is then given by
$$\theta_N=0.314{{r_0 \cos{\gamma}}\over{\bar{h}}}, \eqno(5)$$
where \citep{Fried1982}
$$\bar{h}=\sec{\gamma}\Big(\int{C_n^2(h)h^{5/3}dh}\Big/{\int{C_n^2(h)dh}}\Big)^{-3/5}. \eqno(6)$$
Now appropriately modified,
equations 4 and 5 will give an estimate of the
isoplanatic angle and layer height from measurements of the off-axis Strehl ratio. 
In practice, these Strehl ratio
measurements could be made with a calibration observation of a suitably
dense star field. Unfortunately, the Strehl ratio, both on-axis and off-axis,
will vary with the seeing. In fact, since the off-axis PSF depends on 
not just the quickly varying $r_0$, but also on an evolving $C_n^2$ profile, it 
will vary even more rapidly, and with greater amplitude \citep{Rigaut1999}. 
Thus, if the image is to be obtained as a mosaic, it
must be completed quickly, before changes in $r_0$ or $\bar{h}$ can 
introduce objectionable inhomogeneities.
This calibration measurement, plus knowledge of the on-axis PSF, 
could be used to predict the off-axis PSF as long as there is no significant
variation in the height of the dominant turbulence layer(s) or change in the
equivalent number of compensated modes.

It should be stressed that the off-axis calibration image contains degenerate 
information about $\bar{h}$ and $N$. Furthermore, if one does not know $r_0$ during the on-axis PSF measurement, its value of Strehl ratio is also
degenerate, but in $r_0$ and $N$.  That is, as $r_0$ changes so does
$N$, which in turn affects the isoplanatic angle. 
Although the value of $r_0$ can be
determined from the AO system wavefront-sensor signals, thus relieving
the on-axis PSF ambiguity, an independent measure of $\bar{h}$ is 
more difficult to obtain.  Taking SCIDAR measurements during the AO observations could provide this information, but at substantial cost and effort.  However, even with the
degeneracy one might obtain useful results. If a
uniform and low value of $N$ were maintained, the off-axis PSF would change 
very gradually with offset.  
Figure~\ref{figure_star_onaxis_offaxis} illustrates the case where Strehl ratios are high, and small changes in $r_0$ will
dramatically affect not only the on-axis Strehl ratio, but also the spatial 
variation in the PSF.  A ``first-order'' method of determining the off-axis
PSF variation - one that does not account for the variation in $N$ - may be less successful here.

In response to a need to obtain reliable off-axis PSFs for the processing
of faint galaxies, one of us (ES) developed a
means of providing PSF information at various field positions
using calibration observations of dense star fields \citep{Hutchings1998, Hutchings1999, Steinbring1998, Steinbring2001, Steinbring2002}.
In this method, the observer obtains images
of a dense star field once or at most a few times each night; the core of a
globular cluster works best. 
The AO system is guided on a star of similar brightness to the one used in the
target observations. Frequently, the required PSF is at an off-axis position
outside the FOV of the camera, so a mosaic of the star field is 
generated by repositioning the telescope at successively greater off-axis
distances. The goal is to obtain an
unsaturated image of the guide star as well as a roughly uniform
distribution of stars at many off-axis positions. Now, although
the guide star in the star-field calibration image is chosen to be similar in
brightness to the one used in the target observation, it is very
unlikely that $r_0$ and, thus, the delivered PSF is identical. Therefore, 
the full PSF
reconstruction technique must fold in information about the on-axis PSF 
{\it during} the scientific observation with the star-field mosaic data. 

Our method to deal with time dependence
is to deconvolve the image of the dense star field with the image
of the guide star in that field. We assume that the effects of 
field-dependent aberrations in the camera optics are small compared to anisoplanicity. We would expect, if the
full image
is deconvolved with a sub-image of just the guide star until the entire
flux is contained within one pixel, that re-convolution with that same
sub-image will reconstitute the original image. That is, the PSF at
each of the off-axis star positions will be restored.
Therefore,
we should be able to take the deconvolved image of the star field and
convolve it {\it with the image of the guide star from a target
observation}. The result would be the image of the star field as if
it were observed under the natural seeing conditions at the time
of the target observation. In this way, the image of the deconvolved star field
contains the differences between the on-axis and off-axis PSFs. 
It is a map of the convolution kernel necessary to
restore the off-axis PSFs over the field and will be referred to here
as the PSF kernel map. 
Equations 4 and 5 suggest that the accuracy of predictions with
the kernel map will decrease if the values of $\bar{h}$ and $N$ during construction of the mosaic differ significantly from those during the target observation. To restate this another way, the off-axis PSF cannot be exactly
separated into an on-axis PSF which depends only on $r_0$, and an off-axis
kernel which depends only on $\bar{h}$.
Even so, we propose that the PSF kernel map is a 
useful approximation and that it provides a plausible method for predicting 
the off-axis PSF. It is not a mathematically exact method, and thus the results
may depend on the deconvolution technique, number of iterations, convergence crition, among other things. For example, deconvolution of the guide-star
to a spike of one pixel in width may lead to a spatial-frequency 
finite-support problem. One way to remedy this would be to convolve with
the updated on-axis PSF first, and then deconvolve with the guide star
from the calibration field. 
However, we will show that variation in the atmospheric conditions during observation of the star field can account for most of the uncertainty in the results, and thus for our demonstration the details of the deconvolution are 
not as important.

Observation of the calibration field is difficult and time-consuming, 
and must be a consideration in the application of the method. 
If only a small number of targets are observed, i.e. at only a few offset angles, a
more practical method may be to measure the kernel with binaries of the
appropriate separations. This may also allow for the interleaving of calibration
and scientific observations, which the scarcity of globular clusters makes
more difficult.  We will show, however, that even a single calibration observation taken during the night is helpful.  It may be that the problem
is sufficiently well constrained that if one knows the on-axis Strehl
and the Strehl degradation at one point, a model dependent on $N$
and $\bar{h}$ will be sufficient for reconstruction of the kernel map, but we 
leave that work for a later paper.

A theoretical expression for PSF
anisoplanatism has been derived by \citet{Fusco2000}, but needs an independent
measurement of $C_n^2$ (in their example case, from thermosonde data) to provide
the kernel. In this paper we focus on what results can be obtained 
from just a calibration image. A benefit of a semi-empirical approach is that 
it will apply to all
AO systems.  Conjugation of the AO correction to high altitude layers, such
as employed in Gemini Altair, will reduce variation in the off-axis
PSF, and should improve the results from our method. Multi-conjugate AO (MCAO) should lessen PSF variation even more. In MCAO, PSF uniformity is more dependent on system geometry than atmospheric conditions, and a direct measurement of the PSF kernel map may prove to be a very successful tool. This may allow field dependent deconvolution, for example.
 
Broad-band AO observations of dense star fields are frequently made by 
other observers, but almost always for scientific observations of the stars in those
fields and not for calibration purposes. They are generally unusable for our analysis for one or more of the following reasons: 1. The image of the guide star is saturated. 2. The FOV
of the mosaic is too small or contains too few stars. 3. Typically, a long 
time elapses between the observation of the guide star and the other parts 
of the mosaic. Variations in seeing thus render any measurement of anisoplanicity unreliable.

The work discussed here involves the observation and analysis of dense
star fields for the sole purpose of calibrating the AO off-axis PSF. 
The observations and data reductions are outlined in 
Section~\ref{observations}.  Computer simulations were used to further
characterize the data and the PSF prediction method was applied to
them. These analyses are found in Section~\ref{analysis} along with a 
discussion of the uncertainties in the predictions.  
Our conclusions follow in Section~\ref{conclusions}.  

\section{Observations and Data Reduction}\label{observations}

Observations of dense star fields were made with two AO systems: Pueo on the 
3.6 m CFHT on Mauna
Kea, and the Lawrence Livermore National Labs (LLNL) AO system on the 3.0 m 
Shane Telescope of the Lick Observatory.
Both are compact systems capable only of low-order correction - providing fewer than 100 actuators in their deformable
mirrors. Both are mounted at the Cassegrain focus. 
However, these systems represent two very different AO designs.
The Pueo is a curvature wavefront-sensing system using avalanche 
photo-diodes (APDs) in its sensor and employing a 19-element bimorph mirror for correction \citep{Rigaut1998}. 
The LLNL AO system uses a CCD-based Shack Hartmann sensor and 
a piezostack mirror with 61 active elements \citep{Bauman1999}. 

The Pueo and Lick AO systems are also located at very 
different observing sites.
The CFHT is at an elevation of 4200 m and benefits from the already excellent seeing of Mauna Kea. Also, the CFHT observing floor is refrigerated to 
obtain good dome seeing
by minimizing the temperature
difference between the telescope and outside air. The Shane telescope
depends on passive radiation and is at an elevation of only 1350 m. As a result, the CFHT provides $V$-band median-seeing conditions of FWHM$=0\farcs65$, whereas at the Shane Telescope the median seeing is over 1{\arcsec}.

The image quality delivered by these systems depends
on both atmospheric
conditions and system performance. Guide stars must be selected to have sufficient brightness to operate the wave-front sensors.
In practice, the LLNL system can obtain diffraction-limited performance 
with a star as faint as $V=12$, while the low-noise
APDs of the CFHT system permit diffraction-limited performance for guide stars
as faint as $V=14$. 
Even the selection of a sufficiently bright guide star, however, does
not guarantee consistent performance. 
For example, the encroachment of thin cirrus clouds has been seen to temporarily attenuate guide-star signals and degrade phase measurement. This causes poorer delivered correction. Furthermore, small changes 
in the AO systems themselves also have an effect. Two examples are variation in the setup and calibration of the optical bench in the afternoon before
observing and flexure of the system
due to gravity during the night. It would therefore be a very difficult
task to model {\it in detail} the performance of either system with a computer simulation in order to compare with observations.

A journal of the observations is given in Table~\ref{table_journal_of_observations}. Observations were
made through $H$ or $K_{\rm s}$ filters of globular cluster fields where archival $V$, $R$, or $I$-band Hubble Space Telescope (HST) Wide-Field Planetary Camera 2 (WFPC2) images were already available. The requirement of previous
WFPC2 imaging was adopted for two reasons. First, to ensure that potential
guide stars were not doubles or otherwise too crowded for use with the AO
system. This was impossible
to determine from previous ground-based images. Second, the large 
2.6{\arcmin} $\times$ 2.6{\arcmin} FOV of WFPC2 helped to provide astrometry 
for registration of the mosaiced AO images.

\subsection{CFHT Pueo}\label{cfht_pueo}

\begin{figure}
\plotone{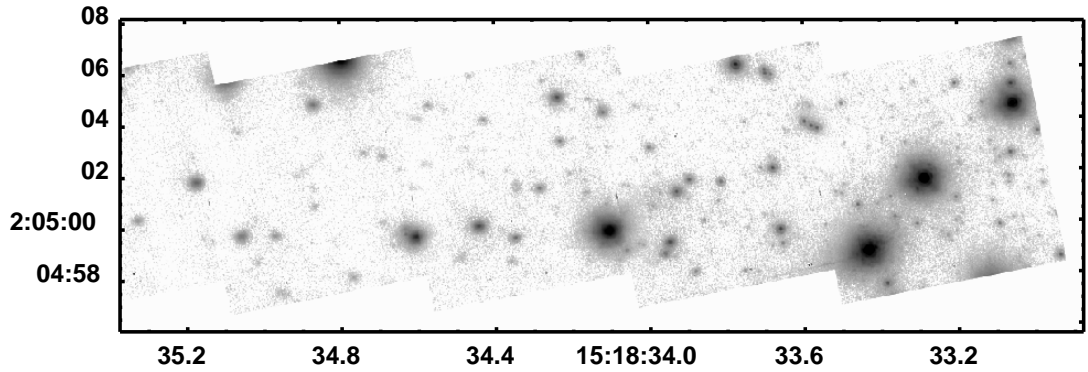}
\caption{A single mosaiced strip from the CFHT Pueo (M 5) dataset. North is
up and east left in this log grey-scale image. Right ascension and declination
are given in J2000.0 coordinates. The guide star
is on the right, the central member of the group of three bright stars.}
\label{figure_m5}
\end{figure}

The CFHT observations were made during the commissioning of Pueo in June 1996.
Observations of the globular cluster \objectname{M 5} were made; two observations using the same $V=10$ guide star were made during one night.
The generation of a wide-field mosaic was made difficult by the small FOV 
(9{\arcsec} $\times$ 9{\arcsec}) of the Montreal Near Infrared 
Camera (MONICA) \citep{Nadeau1994}. A mosaic
pattern was used to obtain a roughly 12{\arcsec} wide by 30{\arcsec} long strip
with the guide-star at one end. The mosaic was composed of 6 individual
pointings of the telescope, with a total exposure time of 30 seconds at each. These images overlapped their neighbours by a few arcseconds, which allowed
us to monitor the seeing by comparing multiple observations of the same
star. Each of the complete mosaics was 
constructed in under 10 minutes. Figure~\ref{figure_m5} is an image of the resulting field. 

The star used for guiding was bright, 4 magnitudes brighter than the faint limit
for Pueo, and sufficient for optimal performance of the system. However, there
was some cirrus present during the night, and this may have caused the delivered
performance to fluctuate independent of seeing variation. 
No independent measurements of the seeing, such as those provided by a dedicated seeing monitor instrument, were available at CFHT during this run.
However, wave-front sensor telemetry was recorded automatically by Pueo.
Values of $r_0(0.5 \mu{\rm m}, {\rm zenith})$ were estimated using the
technique of \citet{Veran1997} by means of software provided on the Pueo website.
Values ranged from 20 to 30 cm, consistent with median or better seeing 
on Mauna Kea.

\subsection{Lick AO}\label{lick_ao} 

Since only two observations were made with the CFHT, they
provided an insufficient dataset for measuring the variation in anisoplanicity.
To measure this we
needed many observations of the same cluster taken over the course of an 
entire night. Such observations were obtained with the Lick AO system 
on the globular cluster \objectname{M 15}, as
part of a program funded by the Center for Adaptive
Optics to study AO PSFs. 

\begin{figure}
\plotone{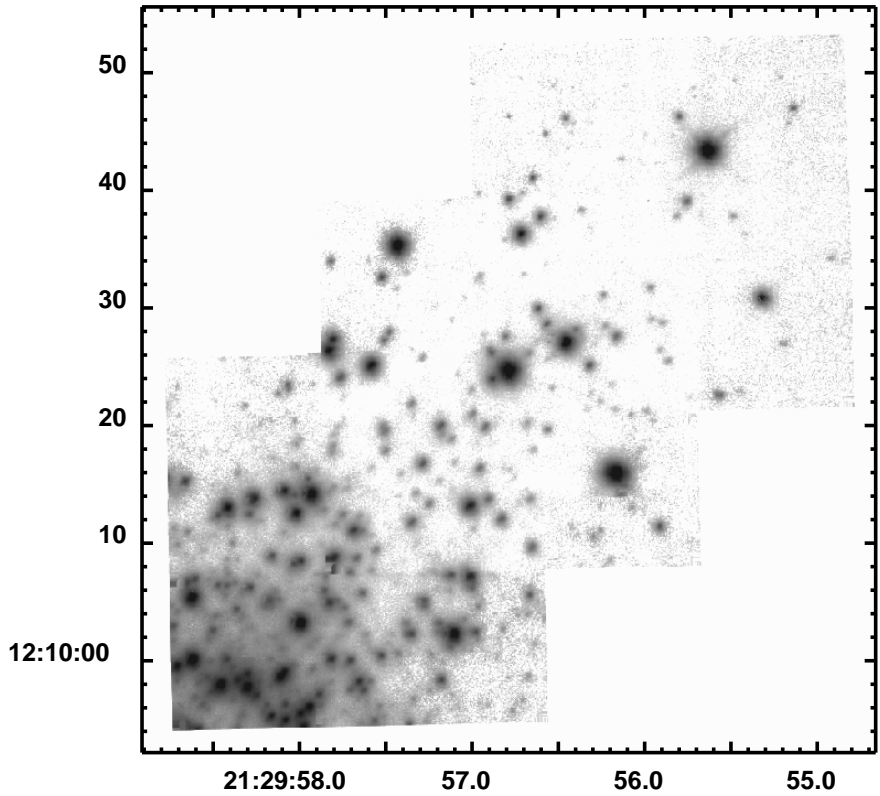}
\caption{A single mosaiced strip from the Lick AO (M 15) dataset. North is
up and east left in this log grey-scale image. Right ascension and declination
are given in J2000.0 coordinates. The guide is
the bright star at upper right. Notice the increased crowding toward the lower
left in the image, which is in the direction of the M 15 core. These stars 
were too crowded for good photometry, which made much of the
data from this region unusable.}
\label{figure_m15}
\end{figure}

Lick AO observations were obtained through a $K_{\rm s}$ filter with
the Infrared Camera 
for Adaptive Optics at Lick (IRCAL) \citep{Lloyd2000} in September 2000. 
The same guide star was used for all star-field observations.
The small IRCAL FOV  (19{\arcsec} $\times$ 19{\arcsec}) required the construction of oblong mosaics
much like that obtained at CFHT. In this case, three individual pointings
of the telescope produced strips roughly 20{\arcsec} wide by 60{\arcsec} long.
Figure~\ref{figure_m15} is an image of the resulting field. 
The exposure time at each pointing was 100 seconds, and each strip
took approximately 9 minutes to complete.
A total of 10 strips was obtained over two nights.  

The guide star was selected
because it was in a particularily dense part of M 15, but it was fainter 
than ideal. It was at the faint limit for Lick AO guide stars. Thus, a nearby but brighter isolated star was also observed from
time to time to track optimal performance. 
We interleaved the mosaics of \objectname{M 15} with observations
of this bright star and obtained total integrations
of 50 seconds with both the AO system providing full correction and
tip-tilt correction only.  

No seeing monitor was available at Lick.
In an effort to obtain independent measures of the seeing during our run, we
obtained simultaneous images of stars with a different telescope nearby on the mountain. 
A measurement obtained from within the same dome would have been ideal, but
was impractical.
We utilized the Nickel 1 m telescope, located roughly 0.5 km west of the Shane. Unfortunately, changes in telescope focus made these data unusable.
We were still able, however, to determine $r_0$ during the run.
Wavefront-sensor telemetry was recorded at intervals by the AO system
and used to determined $r_0(0.5 \mu{\rm m}, {\rm zenith})$ in a manner similar to \citet{Veran1997}. 
The mean values were $8\pm3$ cm for 14 September 2000 and $10\pm3$ cm for the following night.

\subsection{Combined Dataset}\label{combined_dataset}

All the globular-cluster data were compared to HST WFPC2 observations of the
same fields in order to determine the plate scale and orientation of the
cameras. We obtained good relative astrometry of stars, with sub-pixel accuracy.
The data were flat-fielded, and aperture photometry was
carried out on all the sky-subtracted images.  

Strehl ratios were calculated for all the 
stars in the following manner: The diffraction pattern for a centrally obstructed circular
aperture was scaled to the pixel sampling of the camera and to a flux of unity. This image was multiplied by the total flux of each measured star, shifted
to that star's sub-pixel coordinates, and divided by the original sky-subtracted image. The ratio of the peaks gives the Strehl ratio. The shifting of the
diffraction pattern is easily done (in this case, with IMSHIFT in IRAF) 
to sub-pixel accuracy, which will limit the error due to different
pixellation in the two images.  A better method may be to resample the
original image by a factor of 2 or more to reduce the intra-pixel averaging,
but we are mostly interested in PSFs much poorer than the diffraction limit,
where this effect is less severe.  A further, yet systematic error in this measurement arises due to the finite size of the
photometric aperture. For our photometry this corresponds to
roughly a 10{\%} overestimate of Strehl ratio, which is a consistent bias for all measurements. The dense star 
fields selected provided good sampling of the spatial
variation in Strehl ratio. However, photometric
accuracy, and therefore, accuracy of the measurement, dropped for
fainter stars and for smaller separations. We overcame this problem by
selecting from each field a sub-sample of the brightest and most isolated stars.
We maintained this selection throughout the analysis and obtained an
uncertainty in Strehl ratio for any given star in a single frame that was
not much worse than its random photometric error, or on the order of few
percent.
 
The mosaics ideally give an instantaneous measure of the PSF over the field.
Since a time lapse of several minutes occured during the observation of each mosaic, it instead also samples a temporally varying PSF. The overlapping regions
provide an estimate of this variation.
The Strehl ratio of a particular star in both frames typically varied by
15\% for CFHT and 20\% for Lick, but some discontinuities as
large as 50\% occured for Lick. The mean Strehl ratios in the CFHT
and Lick data are approximately 25\% and 5\% respectively, which suggests
the mean {\it absolute} uncertainty in the Strehl ratio at any position
in a mosaic is about 3\%.

The correction delivered by both AO systems when using bright stars was similar.
The mean on-axis Strehl ratio for the CFHT observations of M 5 was 0.24.
This is comparable to the average value of 0.25 obtained for the Lick observations of the bright isolated star in \objectname{M 15}. Note, of
course, that the CFHT images were obtained in $H$ band while the Lick
images were obtained through a $K_{\rm s}$ filter. Since, to first order $r_0\propto\lambda^{6/5}$, observing at redder wavelengths with Lick compensated somewhat for its poorer seeing conditions. The value of 
$r_0(0.5 \mu{\rm m}, {\rm zenith})$ at CFHT was roughly 25 cm, while at
Lick it was approximately 9 cm, or almost 3 times worse. Correcting these 
values of $r_0$ to the observing wavelengths gives $100\pm20$ cm and $53\pm18$ cm at CFHT
(1.6 $\mu{\rm m}$) and Lick (2.2 $\mu{\rm m}$) respectively, indicating
that the effective seeing conditions are comparable, differing by less
than a factor of 2.

We obtained an estimate of the instrumental isoplanatic angle, $\theta_N$, by 
a least-squares fit of equation 4 to the measurements of Strehl ratios 
for stars in the mosaics. The values are given in Table~\ref{table_journal_of_observations} without correction for zenith
angle. The \objectname{M 5} (CFHT)
observations have a mean $\theta_N$(1.6 $\mu{\rm m}$) of 33{\arcsec}. For
\objectname{M 15} (Lick) it was approximately 50{\arcsec} (2.2 $\mu{\rm m}$) 
on average. Combining these values of $\theta_N$ (after correcting for
zenith angle) with measurements of $r_0$ should yield 
estimates of $\bar{h}$ via equation 5. For the observations in \objectname{M 5} the mean
value of $r_0$ was approximately 25 cm, which suggests that $\bar{h}$ was approximately 5 km at CFHT.
Similarily, $r_0\approx9$ cm for the observations of \objectname{M 15} gives
$\bar{h}\approx2$ km at Lick. 

Since the observations of the mosaics and the bright isolated star in the
\objectname{M 15} Lick data were interleaved, we can get a further diagnostic of the Lick AO system performance by comparing the on-axis Strehl ratios of the two observations.
The difference between that obtained with full correction during the mosaics 
and during observations with tip-tilt only correction of the bright isolated star is a measure of how many more 
Zernike modes were corrected in the mosaic observation.
We found that the Strehl ratio improved, on average, from 0.01 to 0.05, a 
factor 
of 5 over tip-tilt correction alone. This suggests that we achieved 
correction of only a few modes beyond tip and tilt \citep{Noll1976}.
This is plausible because the guide star magnitude was near the useful limit
for the Lick system.

\section{Analysis}\label{analysis}

A goal of the observations was to relate gross changes in, say, $r_0$ or 
the number of corrected modes to the accuracy of the semi-empirical PSF prediction. We therefore want to, first, model the performance of the AO systems and
determine what most affects their delivered isoplanatic angles. Next, we will generate the
PSF kernel maps. Finally, we will relate the accuracy of the predictions
based on these maps
to the stability of the atmospheric conditions and performance of the AO system.

\subsection{Computer Modeling}\label{computer_modeling}

A computer simulation was developed to characterize the data. It is a simple
model that includes a single atmospheric layer, 
a circular unobstructed pupil, and ideal wavefront sensing and 
compensation. The simulation
codes were written by one of us (SH) and are based on a
simulation by \citet{Rigaut1994}. Like that code, the wavefront corrugations are due to Kolmogorov turbulence and are contained in
fixed phase screens; but in our case the model of the AO system is much
less sophisticated and only a single phase screen is employed. This 
phase-screen image can be scaled according to Fried's parameter, $r_0$, and the 
height of the atmospheric layer, $h$. The telescope pupil is then superimposed
at a random position in the phase screen, and the aberrations are decomposed 
into Zernike modes via the prescription of \citet{Noll1976}.
The code simulates compensation of the phase error 
by summing only a small subset, $N$, of the measured modes, re-binning this
result at lower resolution to mimic the true number of actuators in the CFHT 
or Lick system, and subtracting the result from the phase screen. The 
delivered Strehl ratio, $S$, is
calculated at several off-axis positions using the extended Marechal approximation on
the residual phase error in the phase screen. 
The correction is done without time delay, and thus the code simulates
open-loop performance of infinite bandwidth.
All the simulation steps
are repeated for hundreds of iterations, and the output is the average change 
in Strehl ratio with offset. The values of $r_0$, $h$, and $N$ can then
be determined by a best fit optimization to these curves.  
Because only $r_0$ is directly specified by the observations 
(see Section \ref{combined_dataset}), we kept $h$ and $N$ as free parameters in the model. We set $N\leq12$, which models perfect correction of low-order modes such as tip, tilt, defocus, astigmatism, coma, and trefoil.

\begin{figure}
\plotonehalf{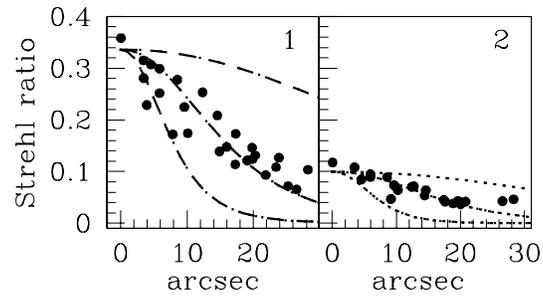}
\caption{Plots of the Strehl ratio as a function of off-axis angle for two observations
of M 5 (CFHT). The filled circles are the data from mosaics 1 and 2. The curves
represent the model results: dotted for 8 corrected modes (mosaic 2) and dot-dashed for 11 corrected modes (mosaic 1). The best-fit models have 
$h=5$ km. Models for 1 km and 10 km are also shown, the latter indicated
by the steepest curve.}
\label{figure_m5_plots}
\end{figure}

Figure~\ref{figure_m5_plots} shows plots of $S$ versus offset for the two
observations at CFHT. The data are shown as filled circles, and
the results of the models are overplotted as smooth curves. The best-fit models have $h\approx5$ km. Poorer fits were obtained with higher and lower altitudes;
the results of $h=1$ km and $h=10$ km are shown. Correction was good for
the first mosaic, suggesting $N=12$, while for the second mosaic $N$ is closer
to 9. For these choices of parameters, the simulations suggest a value of $\theta_N\approx20${\arcsec} for both observations.

\begin{figure}
\plotone{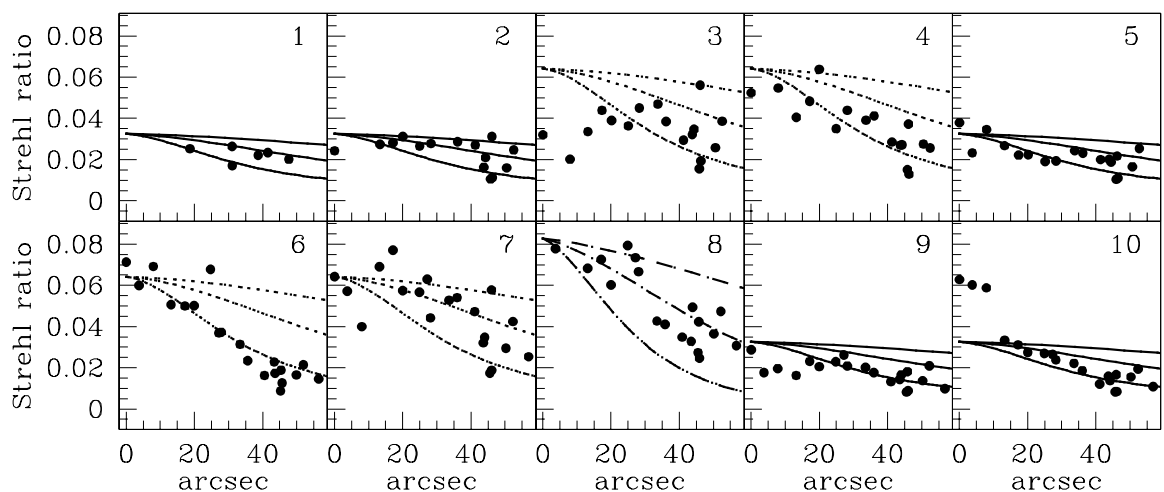}
\caption{Plots of the Strehl ratio as a function of offset for ten observations
of M 15 (Lick).
Mosaics 1 through 5 were obtained sequentially over 5 hours on 14 September
2000, while 6 through 10 were obtained similarly the next night.
Note the strong discontinuities in the data for mosaics 3 and 10. In both
cases this
discontinuity occurs at the boundary between data obtained at two separate
telescope pointings, and thus a plausible explanation is variation of $r_0$ 
with time. The curves
represent the model results: solid for 2 corrected modes, dashed for 3 corrected
modes, and dot-dashed for 4 corrected modes. Each of these is further 
separated into models with atmospheric-layer heights of 1 km, 2 km, and 
4 km; the last indicated by the steepest curve.}
\label{figure_m15_plots}
\end{figure}

Figure~\ref{figure_m15_plots} shows the results for the observations at Lick.
The data are shown as filled circles. The probable cause
of discontinuties in the data for observations 3 and 10 is
variation in $r_0$ during the construction of those mosaics. Several simulations are shown as smooth curves; solid curves represent a simulation with $N=3$,
dotted curves: $N=4$, and dot-dashed curves: $N=5$. Three choices of layer height are plotted for each choice of number of modes corrected. Curves corresponding
to $h$ of 1 km, 2 km, and 4 km are progressively steeper, with $h=4$ km being
the steepest. The models suggest that, for Lick, a low number
of modes were corrected by the system and the dominant layer was at an
altitude between 1 km and 4 km. However, 
The best fit model has $N=3$ for mosaic 1, 2, 5, and 9. A value of
$N=4$ is a better fit for mosaics 3, 4, 6 and 7; and mosaic 8 suggests
$N=5$. To within the
errors of the data the best model-fits all have similar values of $\theta_N$,
tending towards 70 arcseconds.

\subsection{Generating PSF Kernel Maps}\label{generating}

Each of the mosaiced images from both the CFHT and Lick datasets was
deconvolved with the image of its guide star. 
The Lucy-Richardson algorithm was employed. Other deconvolution methods such as
maximum entropy were tried, with similar results. Linear deconvolution
methods may prove to be faster, but were not studied.
Iterations were allowed to continue until
almost all the flux in the deconvolved image of the guide star was contained
within a single pixel. The following convergence criterion
was adopted: iterations ceased when less than 5{\%} of the guide star flux
remained outside the one-pixel spike. This was generally achieved
after a few hundred iterations. Reducing the residual to, say, 1{\%} was much 
more computationally expensive, required thousands of iterations, and 
did not significantly improve the final results.
It should be pointed out that no number of iterations will reduce the residual
to zero. For example, one can imagine that if the centroid of the star is not
centered on a pixel, deconvolution cannot leave a spike that is restricted to
just one pixel. To help remedy this, we applied sub-pixel shifts to the images
so that the guide-star was centered on a pixel.  The residuals for stars
at off-axis positions are all larger than that for the guide-star, and less
sensitive to positional errors.  

This deconvolution process generates the PSF kernel maps discussed in Section~\ref{introduction}. The kernel changes
with increasing offset from the guide star, and the degree of change 
indicates the severity of degradation in the off-axis PSF.
Within a few arcseconds of the guide-star, the kernel can be approximated by a delta-function, but at larger offset it is more Gaussian in shape. 
Offsets between 20{\arcsec} and 40{\arcsec} are of particular interest.
Despite anisoplanicity, AO still provides useful correction at these offsets. 
However, these offsets are larger than the typical FOV of imagers currently available with AO on large telescopes. Thus, simultaneous observation of the guide star and target is not possible, and some means of calibrating the
off-axis PSF is needed. 

\begin{figure}
\plottwo{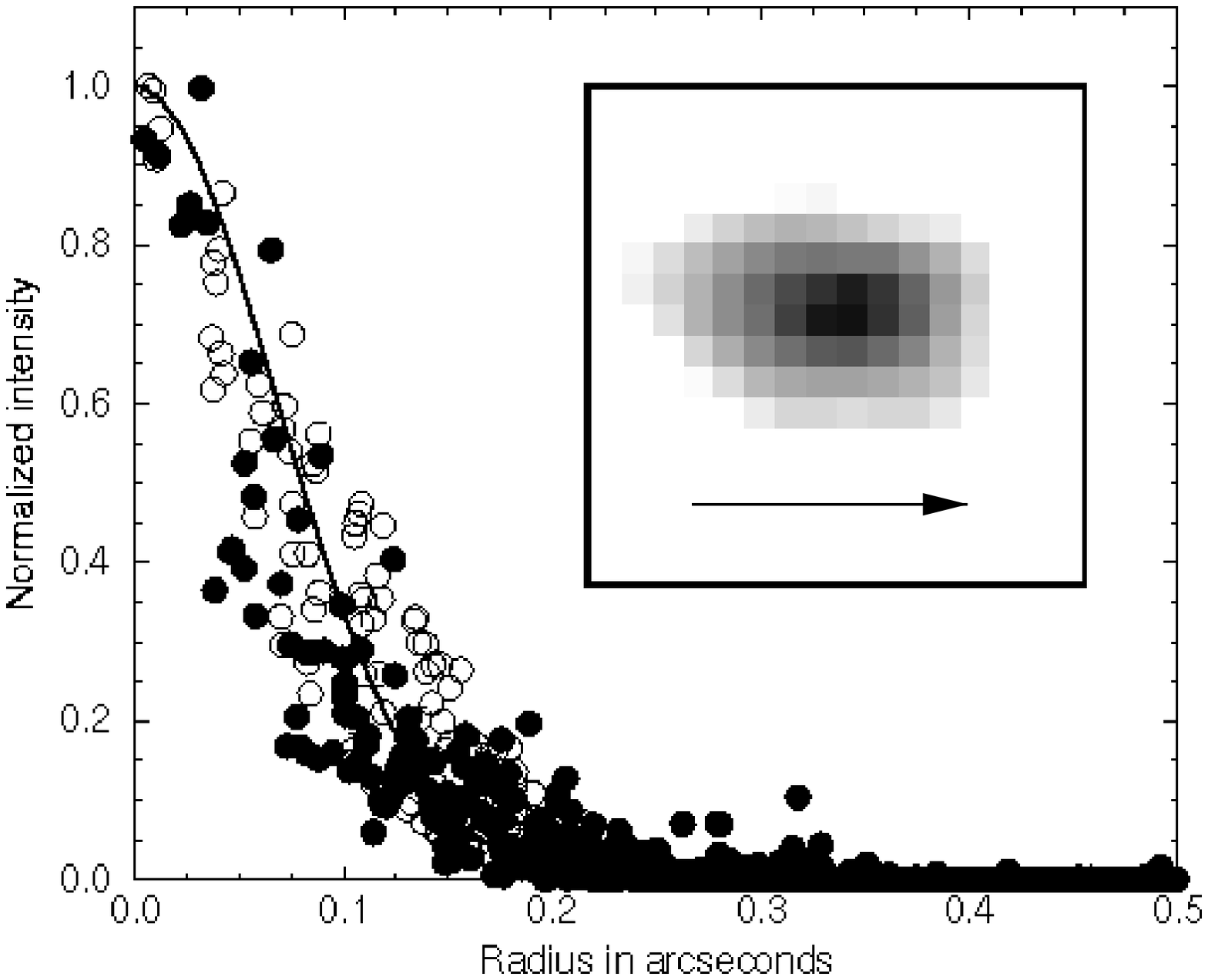}{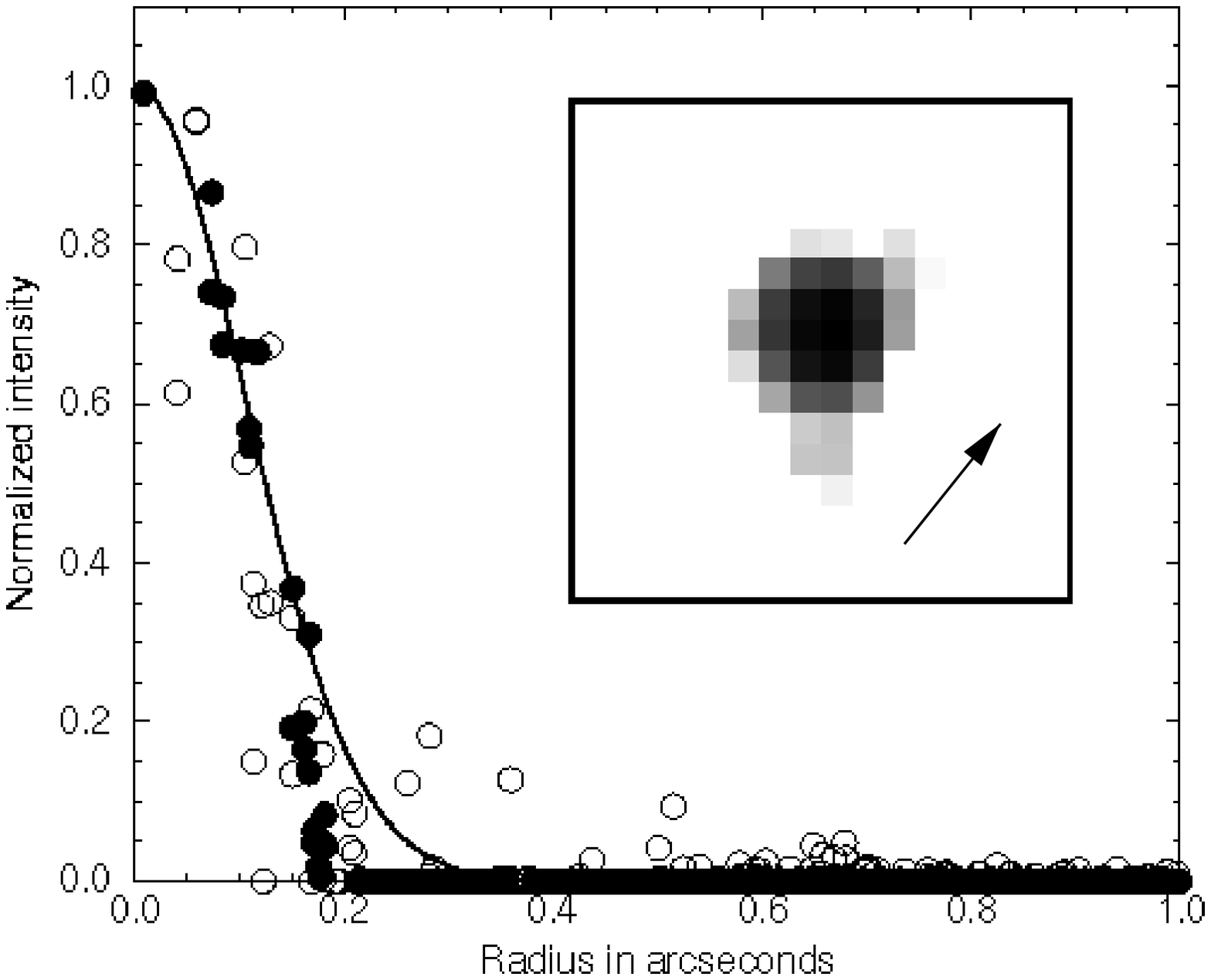}
\caption{Plots of the kernel at an offset of 25{\arcsec} for the
M 5 data (CFHT, left) and M 15 data (Lick, right). Each data point represents
a pixel in the image.
For the M 5 data, the filled circles represent the kernel from the first mosaic, while the open circles represent the second mosaic. For the M 15 data, the filled circles represent the second mosaic and the open circles represent the eighth mosaic. Gaussian fits to the data are overplotted as solid lines.
Two-dimensional representations of the CFHT and Lick kernels are inset.
Each is the average of the same data shown in the plot.
The images have a FOV of
0\farcs5 $\times$ 0\farcs5 (CFHT) and 1{\arcsec} $\times$ 1{\arcsec} (Lick) and are displayed with a log grey-scale. The direction to the guide star is indicated by an arrow. The CFHT and Lick kernels have a different, complex
asymmetric shape.}
\label{figure_m5_m15_kernel}
\end{figure}

Figure~\ref{figure_m5_m15_kernel} shows radial plots 
of kernels for an offset of 25{\arcsec}. 
The kernels for the M 5 (CFHT) data are shown on the left, and two of the ten
kernels for the M 15 (Lick) data are shown on the right. For the CFHT data, the filled circles represent the kernel from the first mosaic, while the open circles represent the second mosaic. For the Lick data, the filled circles represent the second mosaic and the open circles represent the eighth mosaic.
These mosaics
were chosen because they represent the worst-case largest possible difference in on-axis Strehl ratio.
Each datum represents a pixel, and no azimuthal averaging has been done.
One-dimensional Gaussian fits to the data are overplotted for comparison. 
For CFHT, the FHWM of the fit is $0\farcs15$, while for Lick it is $0\farcs25$.
The overlap of the filled and open circles indicates that, for both
the CFHT and Lick kernels, the separate kernel estimates roughly agree.
The scatter in the data is due to variation in $r_0$ during the
observation and to the elongation
of the kernel towards the guide-star, but the true shape of the kernel is somewhat complicated. It is not described by, say, a simple two-dimensional Gaussian with a fixed ellipticity. For example, the wings of the CFHT kernel
can be fit by a Gaussian but the same profile is too broad for the core. The wings of the Lick 
kernel fall off faster than a Gaussian profile. To help illustrate this,
2-dimensional representations of the CFHT and Lick kernels are inset 
in Figure~\ref{figure_m5_m15_kernel}. Each image is an
average of the same data displayed in the plot. Note the complex asymmetric
shapes of both.

We are in the process of parameterizing the variation of the kernel with
field position and have collected more data with Lick AO to assist in that 
work. 
We leave that analysis for a future paper. This paper deals with
the kernel maps in an empirical way and will use them directly to predict the
off-axis PSF.

\subsection{Predicting the Off-Axis PSF}\label{predicting}

A goal of the observations was to determine if the reconstruction
technique outlined in Section~\ref{introduction} could measure
the AO PSF over a large FOV and predict it under different seeing conditions.
Our sample of mosaics consists of long thin strips, and therefore we cannot
use them to measure a change in the PSF with varying position angle about the guide star. 
However, a simple test can check whether a kernel
map generated with different on-axis seeing conditions can be used to find the
first-order
radial variation in the off-axis PSF.
The test is this: reproduce the star field
based on a PSF variation map generated from a different observation of that same
star field.
The best means of demonstrating this is to reconstruct an off-axis PSF
with the use of only an image of the on-axis guide-star and the PSF kernel map.
Then we can compare this reconstructed PSF with the real image of that
star.  For high Strehl ratios, above roughly 30\%, comparison of the
peak brightnesses is the proper figure of merit, because the FWHM will
change little. For lower
Strehl ratios, FWHM becomes more relevant.
Except for the first \objectname{M 5} observation,
the Strehl ratio was 12\% or less for all of the
strips.
Thus, for the most part FWHM and not Strehl ratio is a better
measure of image quality for our data.
We therefore normalized the predicted PSF to have the same peak intensity as the data and compared their FWHM. Our more careful discussion of errors
is left for the \objectname{M 15} data, where Strehl ratio is 10\% or lower.  

\begin{figure}
\plotonehalf{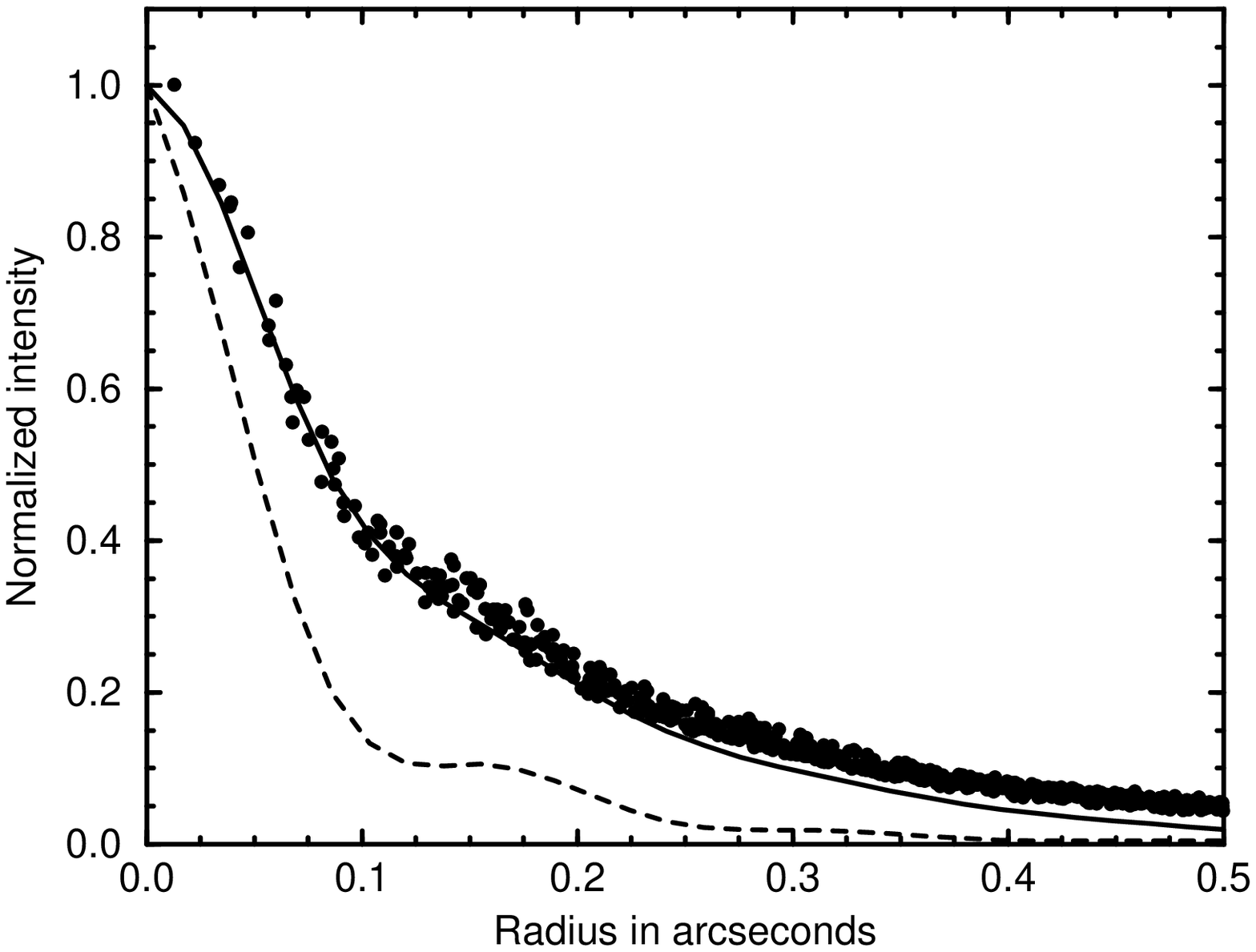}\\
\plotonehalf{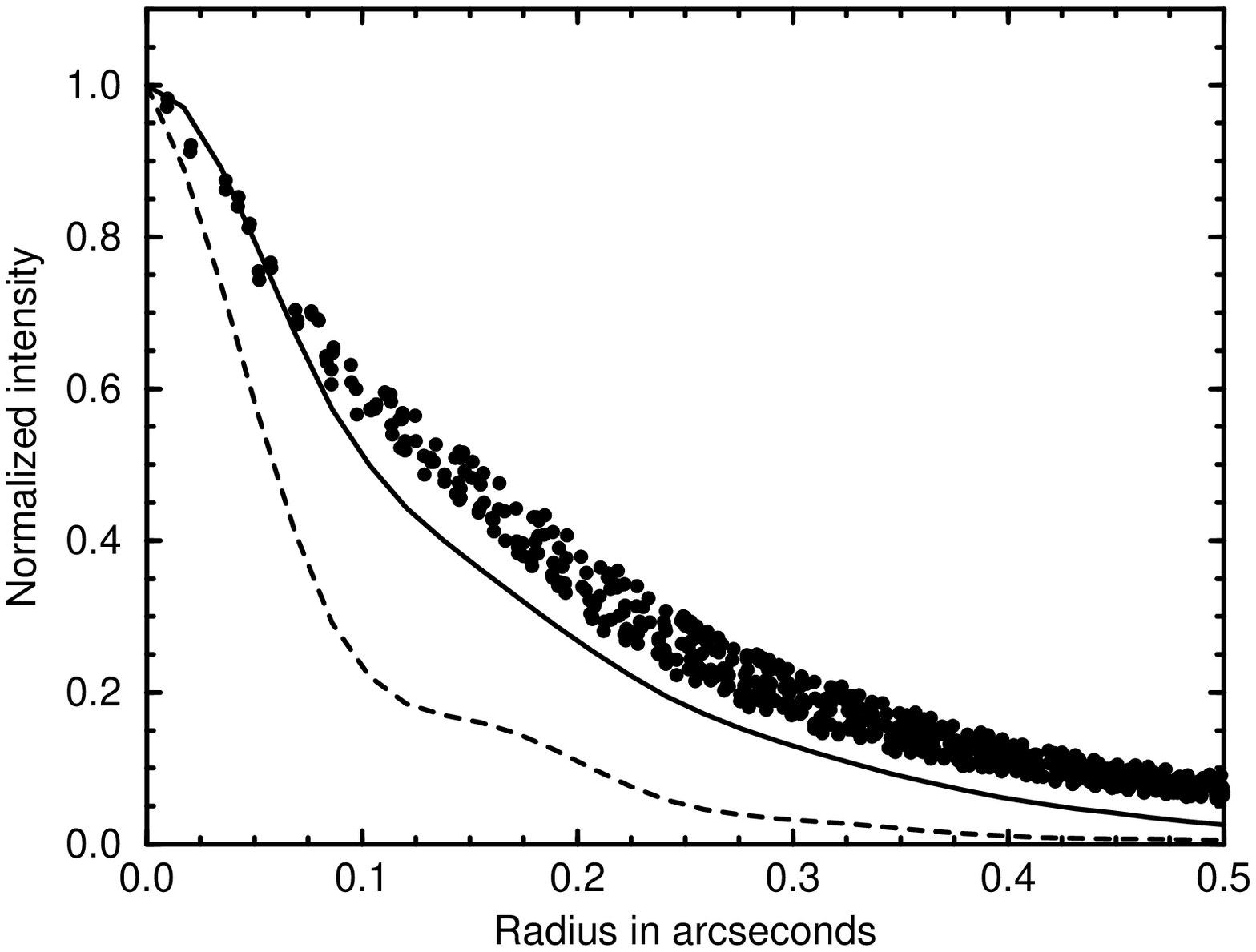}\\
\plotonehalf{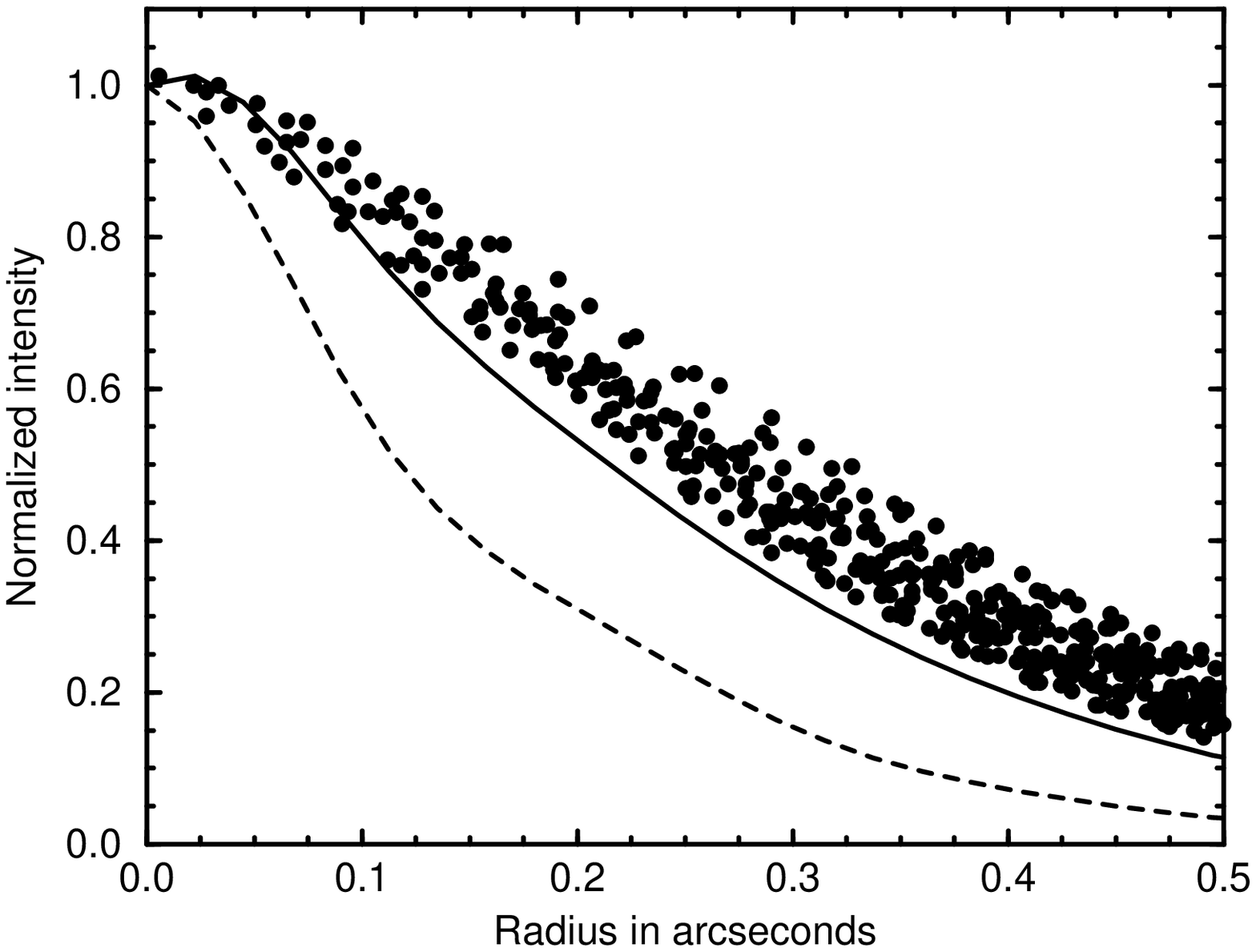}
\caption{Plots of the PSFs in M 5 (CFHT) in mosaic 2 restored with the mosaic 1
kernel map. Shown are offsets of (top to bottom) 0{\arcsec}, 15{\arcsec}, and 25{\arcsec}. The filled circles are the data from
mosaic 2. The dashed curves represent the normalized radial profiles of the PSFs
at the same locations from mosaic 1, and the solid curves represent the model PSFs.}
\label{figure_m5_profiles}
\end{figure}

Figure~\ref{figure_m5_profiles} shows the azimuthally averaged PSF of
the guide star in \objectname{M 5} (CFHT) and similarily averaged PSFs of stars 
at radial distances of approximately 15{\arcsec} and 25{\arcsec}.
The filled circles represent actual data from the second mosaic while
the dashed curves represent the normalized, average PSF profiles from
the first mosaic. The solid
line is the PSF for that location reconstructed from the PSF kernel
using the first mosaic data. The dashed curves are what we would predict the PSF to be at a later point in the night if all that was available was the mosaic from earlier in the night. The model is clearly a better fit to the data than
the earlier PSFs even though it
underpredicts the FWHM. At an offset of 25{\arcsec} the model predicts
a FWHM 31{\%} too small, but without the model the prediction would be
in error by 120{\%}.

\begin{figure}
\plotonehalf{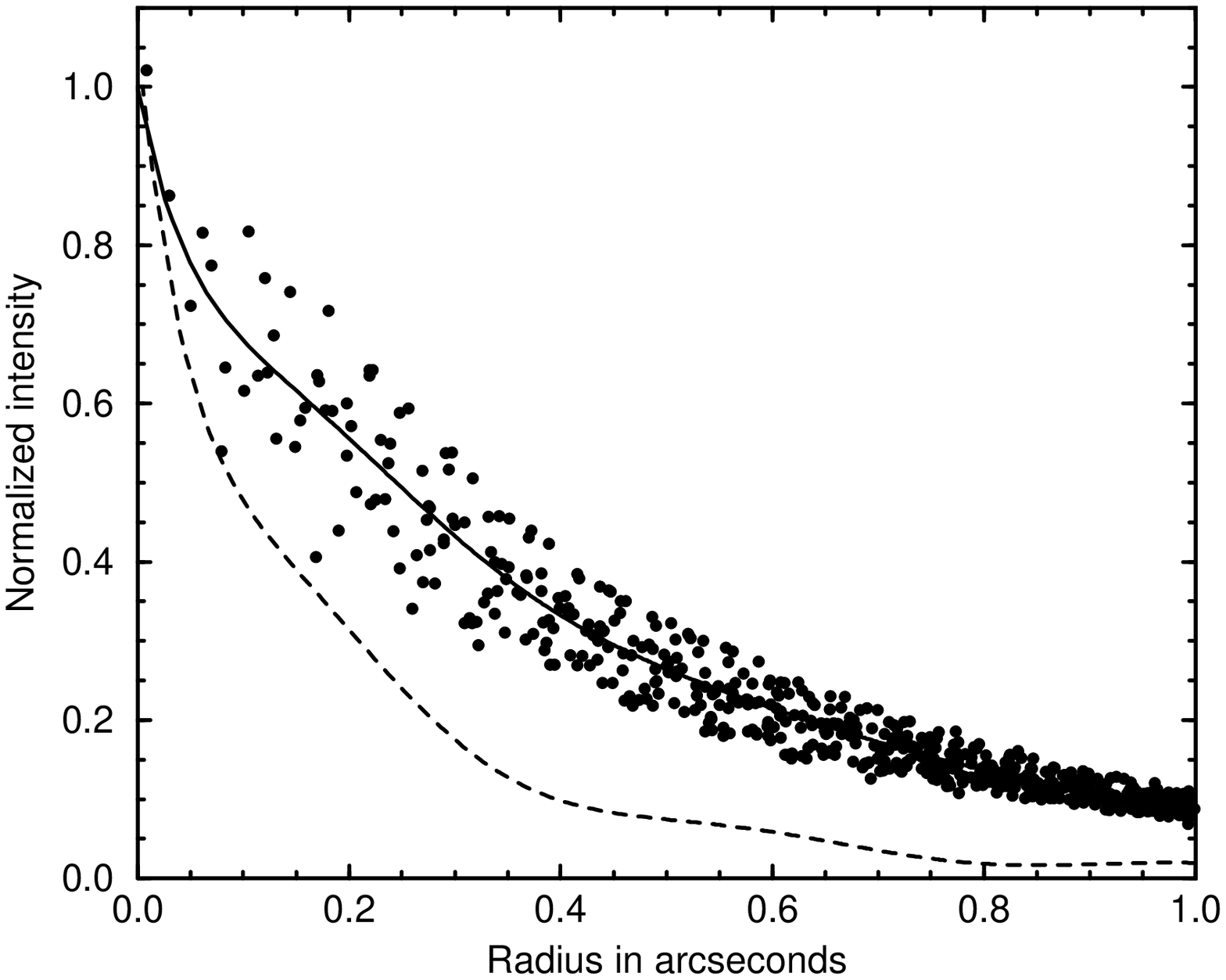}\\
\plotonehalf{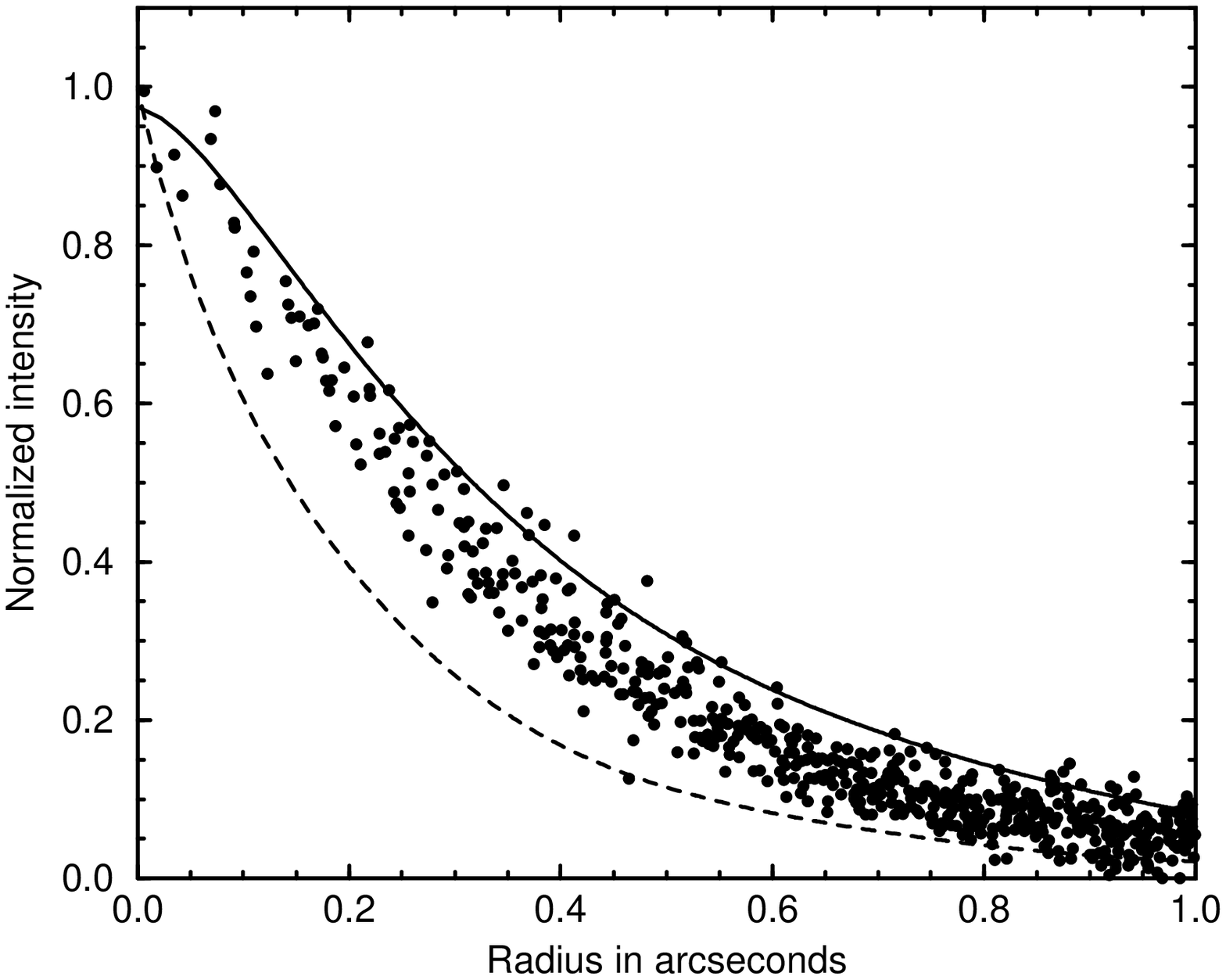}\\
\plotonehalf{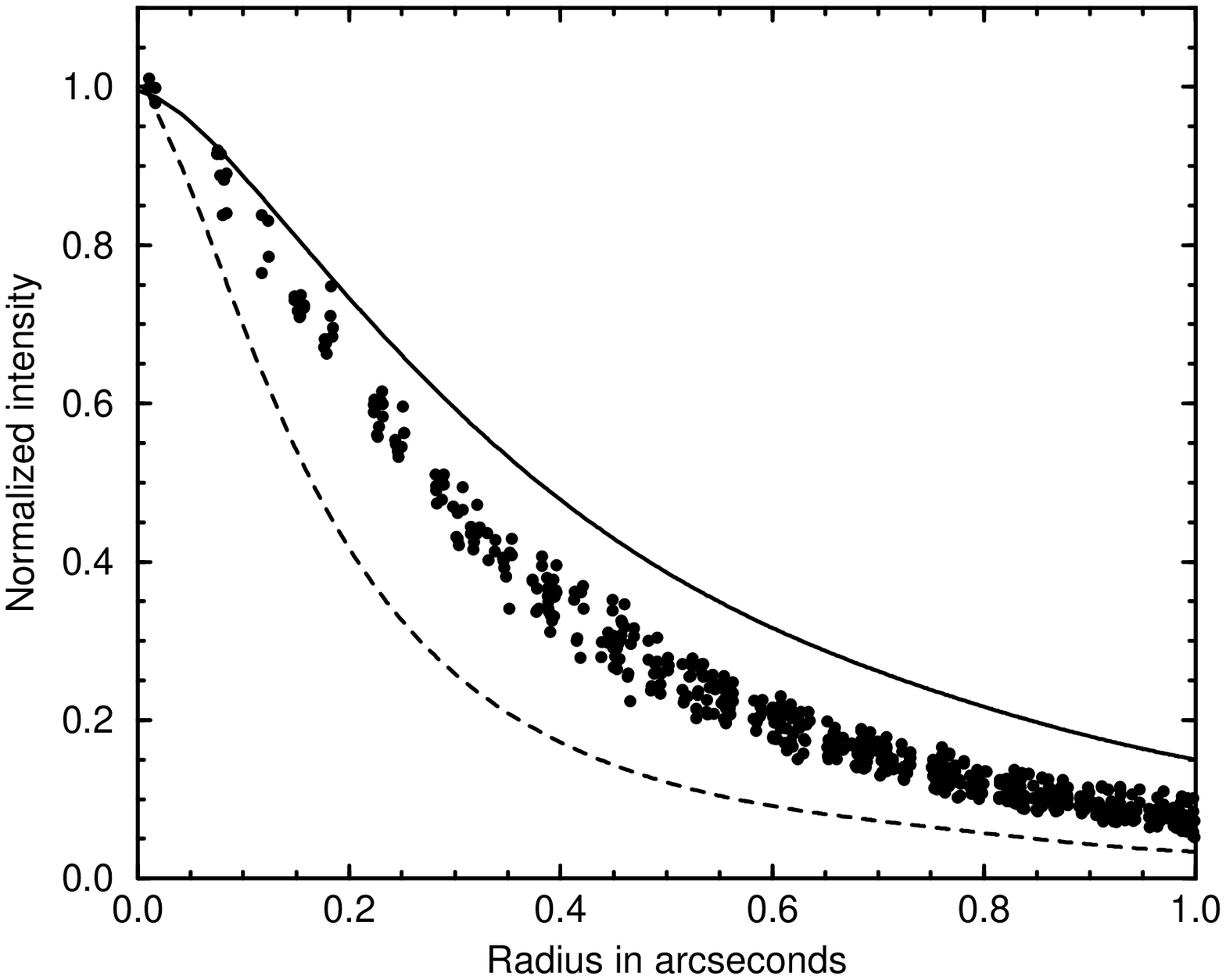}
\caption{Plots of the PSFs in M 15 (Lick) in mosaic 2 restored with the mosaic 8
kernel map. The offsets and symbols are the same as Figure~\ref{figure_m5_profiles}.}
\label{figure_m15_profiles}
\end{figure}

Figure~\ref{figure_m15_profiles} is similar to Figure~\ref{figure_m5_profiles}
but shows the azimuthally averaged PSFs for \objectname{M 15} (Lick). 
Here a kernel map
generated with the eighth \objectname{M 15} mosaic was used to predict the PSFs during the second \objectname{M 15} mosaic observation. These mosaics
were chosen because they represent the worst-case largest possible difference in on-axis Strehl ratio. The results are
qualitatively similar to those for CFHT, but here the
predicted FWHM is broader than the data by 44{\%}. Without the model, the
prediction would be an underestimate and in error by 69{\%}.

\begin{figure}
\plotonehalf{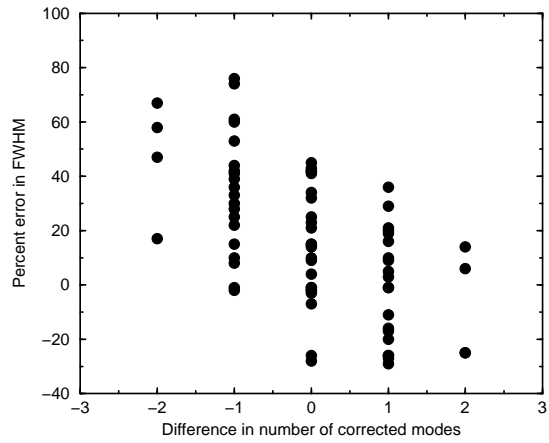}
\caption{Plot of the percent error in predicting the FWHM as a function of the difference in number of corrected modes between the image
used for the kernel map and the predicted image. Each of 
the kernel maps has been used to predict the PSF at an offset of 25{\arcsec} 
in each of the other mosaics. Negative values on the axis correspond to
a smaller number of corrected modes for the mosaic used to generate the
kernel map. This plot suggests that for maximum accuracy the number of corrected modes should be similar, or larger in the kernel mosaic.}
\label{figure_m15_errors}
\end{figure}

The availability of
several mosaics allows us to estimate the uncertainty in predicting the
PSF for the Lick data.  First, each of the guide-star images was used to predict
the PSF at an offset of 25{\arcsec}, i.e. without accounting for anisoplanatism
at all. The FWHM was found to be in error by 60\% on average. The
model was then applied. Each of the images was generated using
each of the other kernel maps.
Figure~\ref{figure_m15_errors} gives results of comparing the observed and predicted FWHM for a star at an offset of 25{\arcsec}. 
The error in predicted FWHM is plotted against the difference in number of corrected modes in the mosaic used for the kernel map versus the 
predicted mosaic. The number of corrected modes was assigned according to the determination in Section~\ref{computer_modeling}.
This plot shows a clear correlation. The trend is towards over-estimating
the FWHM when a mosaic with too few corrected modes is used to predict the
PSF. The mosaics with highest on-axis Strehl ratios, or equivalently, the
highest number of corrected modes were obtained at the lowest airmasses.
This suggests that calibration observations should be taken at equivalent
or lower airmasses than target observations.  
Following this prescription and assuming that an under-estimate and over-estimate of FWHM are equally undesirable, we can determine the mean
error in predicting the PSF. If we had simply used a star at an offset of 25{\arcsec} to predict the PSF, the mean absolute error would be 33{\%}.
For the model PSFs it is 19{\%}; an improvement by a factor of 1.7.
Note that an error of 19{\%} in predicting the PSF is also consistent 
with the 25{\%} uncertainty in Strehl ratio for the Lick mosaics. Thus the
major source of error in prediction of the PSF can be attributed 
to variation in AO correction, and thus plausibly to variation in $r_0$, during construction of the mosaics.  But variation in $r_0$ is not the only way to account for the error. Variation in $N$ and not $r_0$ could arise
from poorer correction due to the presence of cirrus, for example.  This
highlights how difficult it is to make a meaningful empirical prediction,
and that the best method would provide independent measurements of $r_0$, $\bar{h}$, and $N$. The trend in Figure~\ref{figure_m15_errors} suggests 
one path to decreasing the error. The second-order effect of lower angular decorrelation with smaller $N$ could be measured, and an appropriate model
for increasing the off-axis kernel size developed.     
 
\section{Conclusions}\label{conclusions}

We have discussed observations of dense star fields for the purpose of
calibrating AO off-axis PSFs at CFHT and Lick Observatory. The results of our simulations suggest that a simple model with a single
thin turbulence layer is a useful model for our data. For both the 
\objectname{M 5} and \objectname{M 15} observations, a dominant layer at between
2 and 5 km above the telescope and a low number of corrected modes reproduces the anisoplanicity
seen in the data.

A semi-empirical technique was applied that utilizes these findings to improve the prediction of
AO off-axis PSFs. The results show this simple method reduces error
in the prediction 
of the basic radial variation in the PSF.
We have insufficient data to determine
if it can predict the suspected variation in anisoplanicity due to position angle about
the guide star.
The observations of \objectname{M 15} suggest that a single calibration 
observation of a dense stellar field is a useful prediction of anisoplanicity
over the course of a night, or even for a different night, at least for the
atmospheric conditions present at Lick Observatory. Errors in prediction
of FWHM with this method are approximately 19\% compared to 33\%
without any on-axis information, and 60\% without any anisoplanatic correction at all. For maximum accuracy the number of corrected modes should be similar, or larger, in the calibration image.  The accuracy of determining
anisoplanicity from a calibration image is limited 
because the effects of $r_0$, $\bar{h}$, and $N$ are intertwined. Better results could be realized by
separating these effects through independent measurement of $C_n^2$.

The simulations help us understand why the method is successful, if
only modestly. The
dominant-layer height evidently does not change during a night or from night to night
significantly enough to dramatically alter the isoplanatic patch delivered by the system.
Obtaining a calibration image that measures this anisoplanicity is, however, a difficult task. A mosaiced image must be completed quickly,
before variations in delivered Strehl ratio ruin the measurement. Each of our mosaics was constructed in under 10 minutes, but
variation in Strehl ratio by 15\% to 50\% during that time ultimately 
limited the accuracy of the results. 

In September 2001 we again used the Lick AO system but this time employed its laser beacon to acquire more calibration observations of dense star fields.
Instead
of narrow strips, coverage of a large box-shaped region of sky was obtained.
This will allow us to characterize anisoplanatic behavior both radially
away from and azimuthally about the guide star. Future observations would also benefit from confirmatory SCIDAR measurements of the $C_n^2$ profile.   

\acknowledgements

We would like to thank the staffs of the CFHT and Lick Observatories,
and for the helpful comments of the referee, Olivier Lai. 
ES acknowlegdges support from the University of Victoria Department
of Physics and Astronomy. The CFHT observations and some preliminary 
analyses were carried out while he was a graduate student there. 
Portions of this work were performed under the auspices of the U.S. Department
of Energy by the University of California, Lawrence Livermore National 
Laboratory under contract No. W-7405-Eng-48.
This work was supported by the National Science Foundation Science and
Technology Center for Adaptive Optics, managed by the University of California
at Santa Cruz under cooperative agreement No. AST-9876783.

\clearpage

\begin{deluxetable}{lrccccclc}
\tablecolumns{9}
\tablecaption{Journal of Observations\label{table_journal_of_observations}}
\tablewidth{0pt}
\tabletypesize{\small}
\rotate
\tablehead{&\colhead{} &\colhead{} &\colhead{} &\colhead{} &\colhead{} &\multicolumn{3}{c}{AO Correction} \\
\cline{7-9}\\
&\colhead{} &\colhead{Guide} &\colhead{} &\colhead{} &\colhead{} &\multicolumn{2}{c}{Full} &\colhead{Tip-Tilt} \\
\cline{7-8}\\
\colhead{Object} &\colhead{UT} &\colhead{$V$ (mag)} &\colhead{Exp. (s)} &\colhead{Air.} &\colhead{$r_0(0.5\mu{\rm m}, 0^\circ)$} 
&\colhead{$S$ (on-axis)} &\colhead{$\theta_N$ (arcsec)} 
&\colhead{$S$ (on-axis)}}
\startdata
\sidehead{CFHT Pueo/MONICA $H$}
\objectname{M 5}, 1 &25/06/1996 06:06 &10 &\phn30 &1.12 &$30\pm5$ 
&0.36 &\phn$33\pm2$ &\nodata \\
\objectname{M 5}, 2 &\phm{25/06/1996} 07:26 &10 &\phn30 &1.05 &$20\pm5$
&0.12 &\phn$34\pm5$ &\nodata \\
\sidehead{Lick AO/IRCAL $K_{\rm s}$}
\objectname{M 15} PSF &14/09/2000 04:07 &\phn8 &\phn50 &1.26 &\phn$8\pm3$
&0.19 &\nodata &0.01 \\ 
\objectname{M 15}, 1 &\phm{14/09/2000} 04:55 &12 &100 &1.15 &
&\nodata &\phn$82\pm59$ &\nodata \\ 
\objectname{M 15}, 2 &\phm{14/09/2000} 05:47 &12 &100 &1.11 &
&0.02 &\phn$53\pm16$ &\nodata \\ 
\objectname{M 15} PSF &\phm{14/09/2000} 06:28 &\phn8 &\phn50 &1.11 &
&0.22 &\nodata &0.01 \\ 
\objectname{M 15}, 3 &\phm{14/09/2000} 07:11 &12 &100 &1.14 &
&0.03 &\phn$53\pm13$ &\nodata \\ 
\objectname{M 15}, 4 &\phm{14/09/2000} 08:03 &12 &100 &1.24 & 
&0.05 &\phn$47\pm9$ &\nodata \\ 
\objectname{M 15} PSF &\phm{14/09/2000} 08:42 &\phn8 &\phn50 &1.37 &
&0.18 &\nodata &0.01 \\ 
\objectname{M 15}, 5 &\phm{14/09/2000} 09:55 &12 &100 &1.83 &
&0.04 &\phn$56\pm14$ &\nodata \\ 
\objectname{M 15} PSF &15/09/2000 04:18 &\phn8 &\phn50 &1.20 &$10\pm3$
&0.12 &\nodata &0.01 \\ 
\objectname{M 15}, 6 &\phm{15/09/2000} 05:00 &12 &100 &1.14 &
&0.07 &\phn$42\pm6$ &\nodata \\ 
\objectname{M 15}, 7 &\phm{15/09/2000} 05:51 &12 &100 &1.11 &
&0.06 &\phn$48\pm9$ &\nodata \\ 
\objectname{M 15} PSF &\phm{15/09/2000} 06:25 &\phn8 &\phn50 &1.11 &
&0.31 &\nodata &0.01 \\ 
\objectname{M 15}, 8 &\phm{15/09/2000} 07:10 &12 &100 &1.14 &
&0.08 &\phn$46\pm9$ &\nodata \\ 
\objectname{M 15}, 9 &\phm{15/09/2000} 08:05 &12 &100 &1.26 &
&0.03 &\phn$65\pm17$ &\nodata \\ 
\objectname{M 15} PSF &\phm{15/09/2000} 08:43 &\phn8 &\phn50 &1.40 &
&0.50 &\nodata &0.03 \\ 
\objectname{M 15}, 10 &\phm{15/09/2000} 09:53 &12 &100 &1.81 &
&0.06 &\phn$53\pm10$ &\nodata \\ 
\enddata
\end{deluxetable}

\clearpage

\clearpage

\end{document}